\documentclass[12pt]{article}

\usepackage{times}
\usepackage{amssymb}
\usepackage{subfigure}
\usepackage{epsfig}
\usepackage{amsmath}

\newenvironment{sciabstract}{%
\begin{quote} \bf}
{\end{quote}}

\title{Endothermic singlet fission does not proceed via an excimer intermediate}

\author
{Cameron B. Dover,$^{1\dagger}$ Joseph K. Gallaher,$^{1\dagger}$ Laszlo Frazer,$^{1}$ Anthony J. Petty II,$^{2}$\\ Maxwell J. Crossley,$^3$ John E. Anthony$^{2}$ and Timothy W. Schmidt$^{1\ast}$\\
\\
\normalsize{$^{1}$ARC Centre of Excellence in Exciton Science, School of Chemistry,}\\
\normalsize{UNSW Sydney, NSW 2052, Australia}\\
\normalsize{$^{2}$Department of Chemistry, University of Kentucky,}\\
\normalsize{Lexington, Kentucky 40506, United States}\\
\normalsize{$^{3}$School of Chemistry,}\\
\normalsize{The University of Sydney, NSW 2006, Australia}\\
\\
\normalsize{$^\ast$To whom correspondence should be addressed; E-mail:  timothy.schmidt@unsw.edu.au.} \\ \normalsize{$^\dagger$These authors contributed equally to this work.}
}

\date{}

\begin{document}


\baselineskip24pt


\maketitle


\begin{sciabstract}
Singlet fission is a process whereby two triplet excitons can be produced from one photon, potentially increasing the efficiency of photovoltaic devices. Endothermic singlet fission is desired for maximum energy conversion efficiency, and such systems have been shown to form an excimer-like state with multi-excitonic character prior to the appearance of triplets. However, the role of the excimer as an intermediate has, until now, been unclear. Here we show, using 5,12-bis((triisopropylsilyl)ethynyl)tetracene in solution as a prototypical example, that, rather than acting as an intermediate, the excimer serves to trap excited states, to the detriment of singlet fission yield. We clearly demonstrate that singlet fission and its conjugate process, triplet-triplet annihilation, occur at a longer intermolecular distance than an excimer intermediate would impute. These results establish that an endothermic singlet fission material must be designed that avoids excimer formation, thus allowing singlet fission to reach its full potential in enhancing photovoltaic energy conversion.
\end{sciabstract}


Singlet fission (SF) is a process whereby a photo-generated singlet exciton splits into two spin-correlated triplet excitons.\cite{Smith2013a,Smith2010} This can occur where the triplet exciton is about half the energy of the singlet exciton, and there is suitable coupling between two chromophores.\cite{Chan2011,Walker2013,Yost2014,Zimmerman2010,Zirzlmeier2015,Pensack2016} It is of current interest since it offers the possibility to circumvent the Shockley-Queisser limit of single-threshold solar cells, such as those based on crystalline silicon (c-Si), but its detailed mechanism is a matter of some debate.\cite{Zimmerman2010,Zimmerman2011,Chan2012,Walker2013,Feng2016a,Burdett2012,Monahan2015,Bakulin2015,Musser2015,Busby2015,Monahan2016} Exothermic SF, where the energy of two triplet excitons is lower than the initial singlet exciton, has been shown to proceed rapidly, with high yields, and has been incorporated into prototypical devices.\cite{Ehrler2012,Ehrler2012a,Congreve2013,Tritsch2013}
However, the resulting triplet states are too low in energy to be coupled to existing efficient photovoltaic cells. 
Indeed, it has been shown that the most efficient SF solar cell should exhibit fission that is \emph{endothermic}, energetically ``uphill'', corresponding to an upper energy conversion efficiency limit of 45.9\%.\cite{Tayebjee2015,Tayebjee2012} In principle, SF is applicable to any semiconductor material, be it c-Si, perovskite, or chalcogenide.

One class of chromophores that exhibits endothermic fission is the tetracene derivatives. These are especially interesting for potential solar energy applications because their triplet energies ($\sim1.2$\,eV) are slightly above the band gap of silicon (1.12\,eV), offering the possibility to significantly boost the efficiency of c-Si cells. But, the exciton dynamics of tetracene are far from straightforward.\cite{Burdett2012,Tayebjee2013,Wilson2013,Arias2016,Piland2015}

\begin{figure}
\centering
\includegraphics[width=8.5cm]{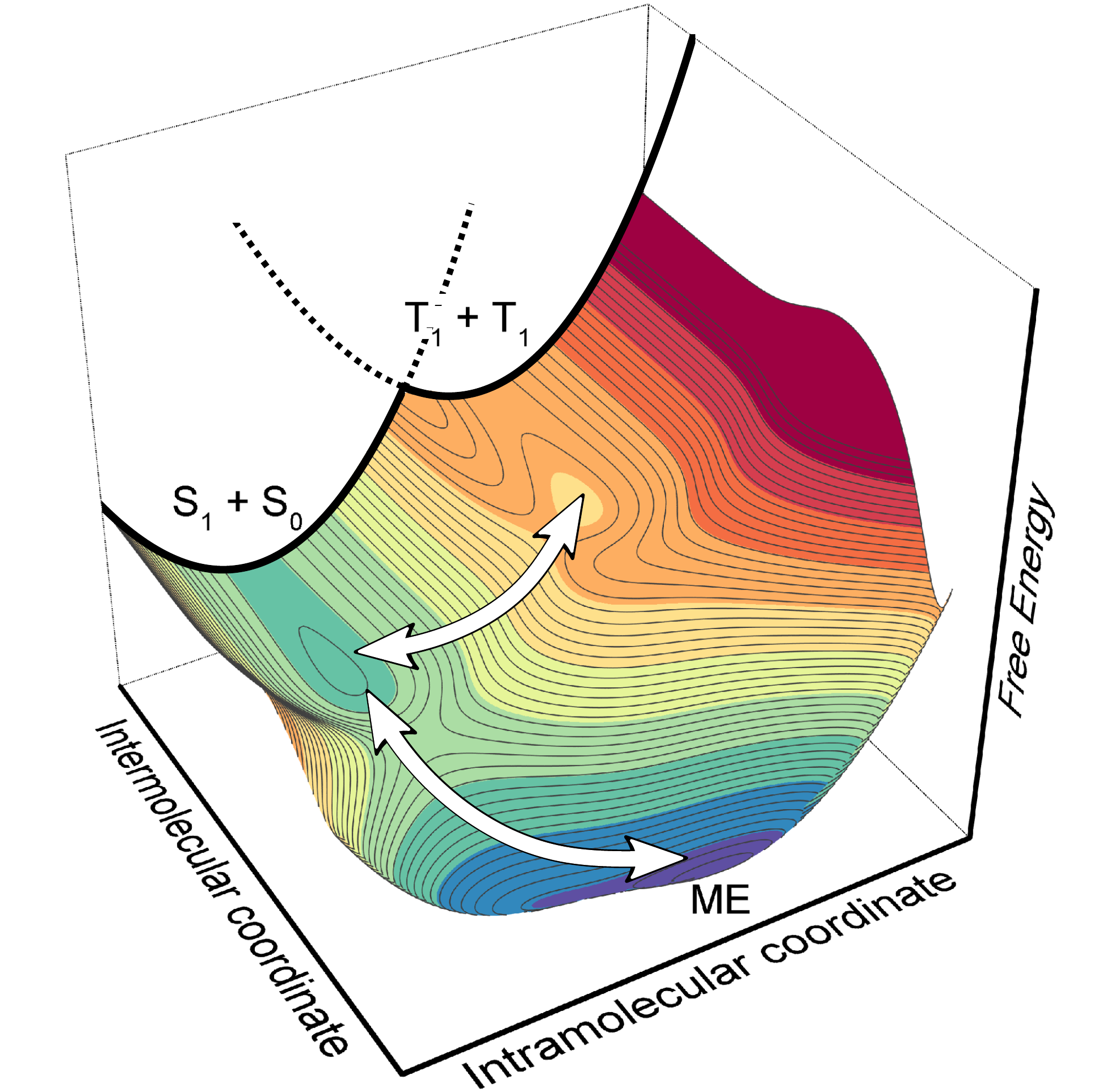}
\caption{Schematic Marcus-Hush-Morse free energy surface of the SF-TTA process. The intramolecular coordinate interconverts singlet and triplet chromophore geometries. The intermolecular coordinate represents the distance between two chromophores. See Supplementary Materials for details.\label{wow}}
\end{figure}

Tetracene films exhibit a rapid dimming of photo-generated singlet excitons with a time-constant of about 80 ps.\cite{Tayebjee2013,Lim2004,Wilson2013} However, while this has been associated with SF, it has been demonstrated to have no significant temperature dependence, despite its endothermic nature. This has necessitated the invocation of a lower energy, spectroscopically dim intermediate state.\cite{Burdett2011,Tayebjee2013} At low temperatures, this ``intermediate'' is trapped, and results in bathochromically shifted excimer-like emission.\cite{Burdett2011,Tayebjee2013}

The idea that endothermic emission proceeds via an excimer of multi-excitonic (ME) character is supported by transient absorption spectroscopy in high concentration solutions of 5,12-bis((triisopropylsilyl)ethynyl)tetracene (TIPS-Tc). Friend and co-workers\cite{Stern2015} observed that the ME excimer exhibits a transient absorption spectrum much closer in appearance to the triplet state than the excited singlet, implying its role as an intermediate in SF. This idea has also been invoked by Mauck \textit{et al.}, who identified excimer-like precursors to SF in thin films of diketopyrrolopyrrole using transient optical spectroscopy.\cite{Mauck2016} These reports clash with a recent theoretical study on the electronic coupling between tetracene motifs predicting that excimer geometries have increased coupling to the ground state and increased rate of radiationless relaxation; therefore, the formation of an ME excimer may in fact be detrimental to SF.\cite{Feng2016b,Korovina2016}

If the ME excimer were an intermediate to SF, it must also be observed in the reverse process, triplet-triplet annihilation.\cite{Parker1962} In this report, we clearly demonstrate that this is not the case, and that a model of the time-resolved photoluminescence is not consistent with an ME excimer intermediate. Rather, SF primarily results from a direct pathway with the ME excimer acting as a trap, as illustrated in Figure \ref{wow}. This has important implications for the design of endothermic SF materials:  ME excimer formation is to be avoided, as it does not aid in the SF process, but rather hinders it.

\begin{figure}
\centering
\includegraphics[width=8.5cm]{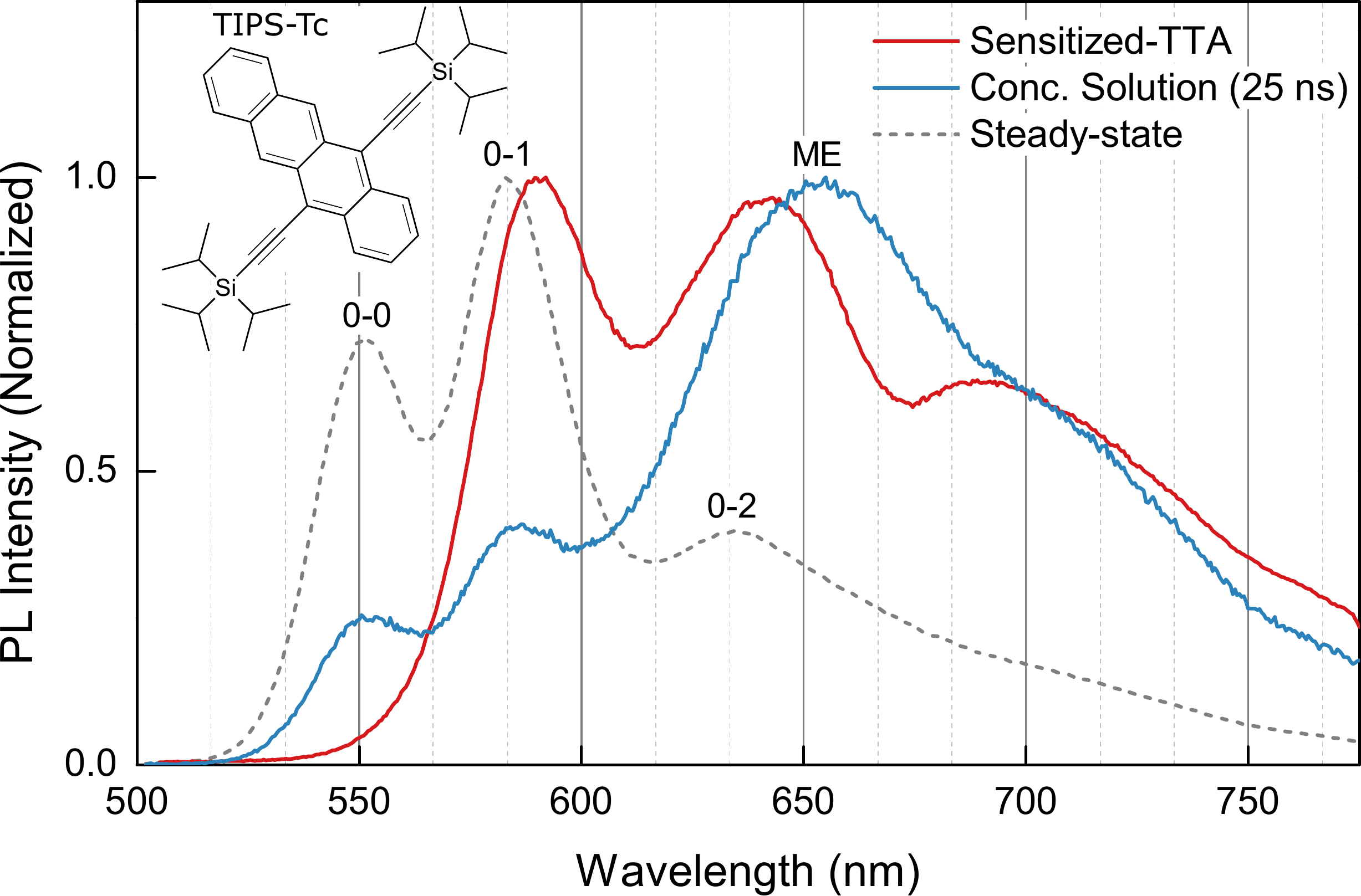}
\caption{Emission spectra of concentrated TIPS-Tc solutions. The sensitized TTA spectrum is influenced by the absorption spectrum of the sensitizer (PdPQ$_4$ - see supplementary materials)\label{TTA}}
\end{figure}

\begin{figure}
\centering
\includegraphics[width=8.5cm]{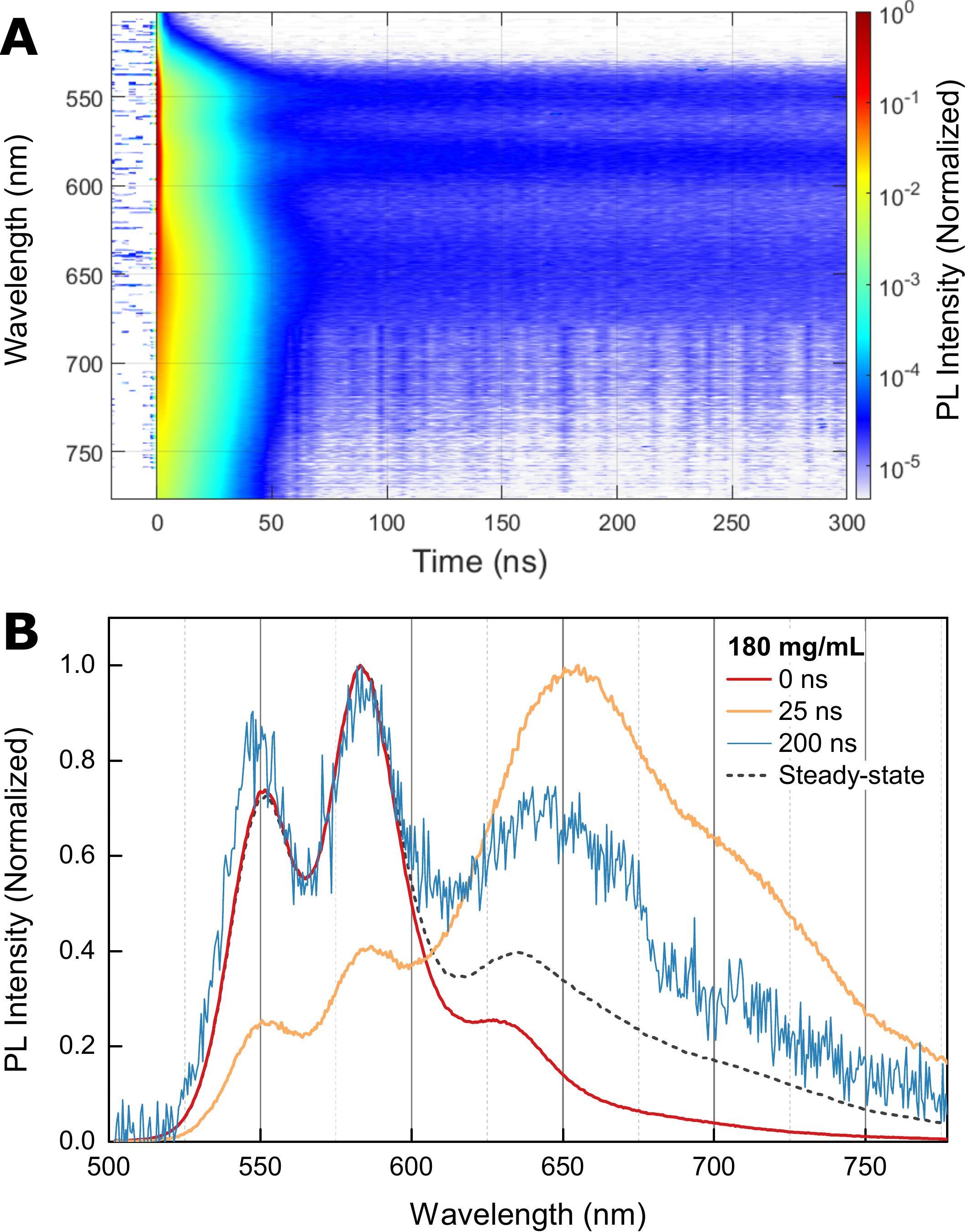}
\caption{A) Time-resolved photoluminescence of concentrated TIPS-Tc in toluene. B) Spectral slices illustrating the spectral dynamics.\label{TRPL}}
\end{figure}

Figure \ref{TTA} (red trace) displays the time-integrated photoluminescence resulting from sensitized triplet-triplet annihilation (TTA) in a concentrated TIPS-Tc solution (180\,mg/mL). The most intense peak corresponds to the $0-1$ emission band of the $S_1$ state of TIPS-Tc, the $0-0$ band having been suppressed due to self-absorption and by the triplet sensitizer, a palladium porphyrin (see supplementary materials for modelling of reabsorption). As the sample was irradiated by a 670\,nm laser pulse, the upconverted TIPS-Tc photoluminescence at shorter wavelengths results from annihilation of TIPS-Tc triplets ($T_1$). In photoluminescence studies of concentrated TIPS-Tc solutions, in which the $S_1$ state is directly excited at 500 nm, the emission rapidly takes on the appearance of an excimer, which has been shown by Friend and co-workers to be of multi-excitonic (ME) character.\cite{Stern2015} The emission spectrum of a concentrated TIPS-Tc solution, 25\,ns after excitation, is shown as the blue trace in Figure \ref{TTA}. The disparity between these spectra (red and blue traces in Figure \ref{TTA}) is striking, and shows that the TTA process, which is the conjugate of SF, does not efficiently regenerate the ME excimer state. To see if this is also true of SF-generated triplets, we performed a time-resolved photoluminescence study of a concentrated TIPS-Tc solution, the results of which are plotted in Figure \ref{TRPL}.

A heat map of the time-resolved photoluminescence (TRPL) as a function of wavelength and time is displayed in panel A of Figure \ref{TRPL}. At very early times, the emission appear to be dominated by free $S_1$ state TIPS-Tc molecules, as shown in panel B (red trace). This rapidly transmutes into an ME-excimer-dominated spectrum which persists out to 50\,ns after laser excitation. The 25\,ns spectrum, shown in panel B, clearly exhibits residual $S_1$ emission evidenced by the $0-0$ and $0-1$ emission bands. At times longer than 50\,ns, the emission again becomes dominated by $S_1$ emission. This is consistent with the long-time emission being due to TTA which, as shown by the sensitized TTA experiments, generates a spectrum dominated by $S_1$ emission. An understanding of the interplay between the various states in this system necessitates a kinetic model.

The photogenerated $S_1$ states pair with ground state $S_0$ molecules to generate ME excimers on a picosecond time scale,
\begin{equation}
S_1 + S_0 \begin{array}{c}
            k_1 \\
            \rightleftharpoons\\
            k_1'
          \end{array}
 ME.
\end{equation}
The ME excimers are widely considered to dissociate to yield free triplets,
\begin{equation}
ME \begin{array}{c}
            k_2 \\
            \rightleftharpoons\\
            k_2'
          \end{array}
 T_1 + T_1.
\end{equation}
Thirdly, as shown by our TTA experiments, there must exist a channel directly linking the free triplets with the $S_1$ manifold,
\begin{equation}
 T_1 + T_1 \begin{array}{c}
            k_3 \\
            \rightleftharpoons\\
            k_3'
          \end{array}
S_1 + S_0.
\end{equation}

In the above processes, to satisfy detailed balance, each reverse process must balance the forward process such that the change in Gibbs energy is $\Delta G_i = -k_B\log(k_i/k_i')$. The cyclic nature of the above processes constrains the overall changes in Gibbs energy such that $\Delta G_1 + \Delta G_2 + \Delta G_3 = 0$. The states $S_1$, ME and $T_1$ decay (non)-radiatively to $S_0$ with rate constants of $k_S$, $k_{ME}$ and $k_T$.

\begin{figure}
\centering
\includegraphics[width=8.5cm]{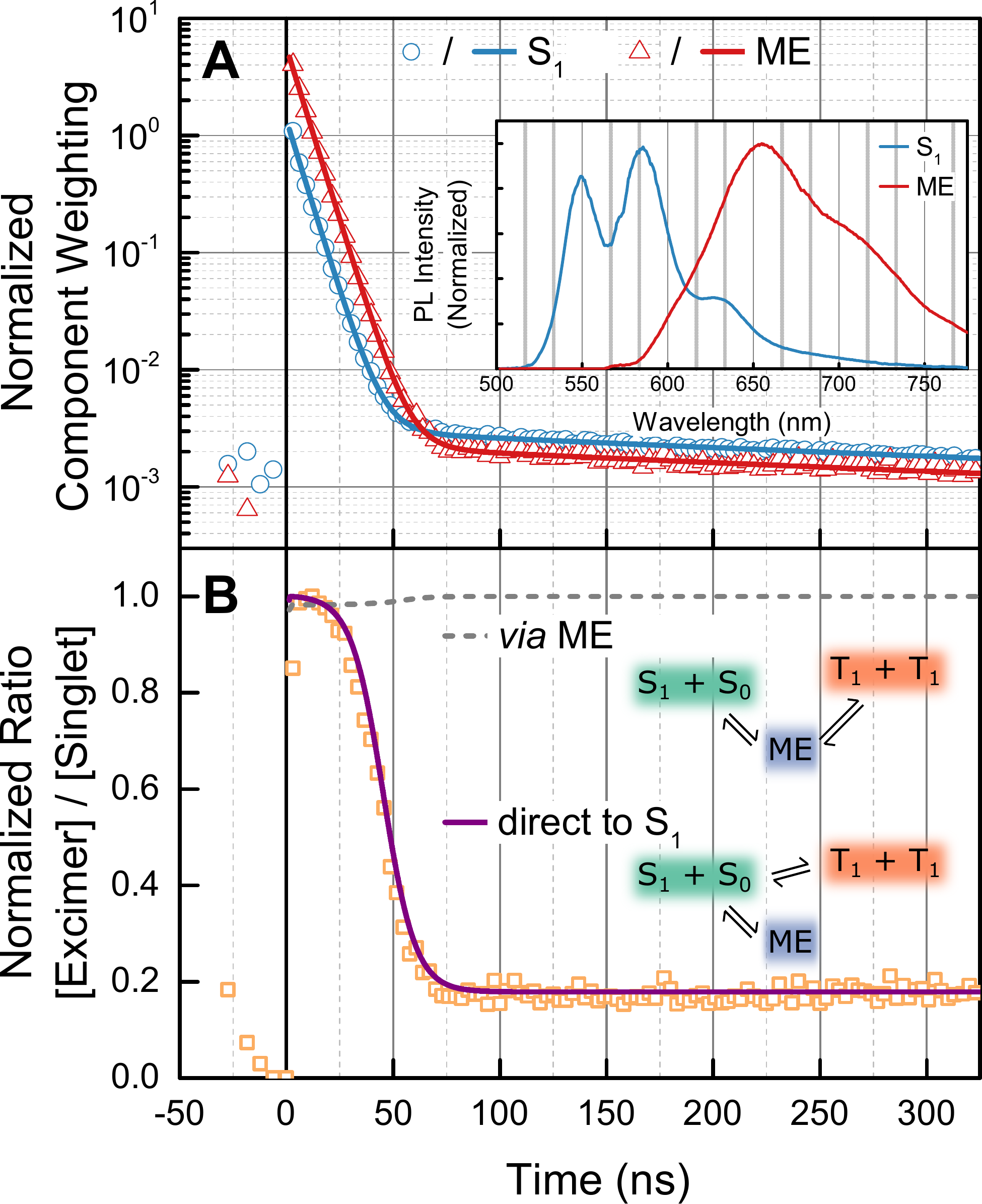}
\caption{A) MCR-ALS extracted $S_1$ and ME excimer emission components, as a function of time (inset: spectral components). B) The evolution of the ME:$S_1$ ratio with time. The grey dashed line is the result of a fit with a model that only incorporates fission via the ME state (see supplementary materials). The solid lines are the results of a fit with a model that only incorporates direct fission from $S_1$. The time-zero data point has been cropped as the time-resolution of the TRPL (3 ns) is insufficient to resolve the prompt initial decay of the $S_1$ population.\label{kinetix}}
\end{figure}

The TRPL in Figure \ref{TRPL} was fit with two independent spectra, with time-varying amplitudes, to extract the time-dependence of the $S_1$ and ME concentrations. The results of this procedure are shown in panel A of Figure \ref{kinetix}. It is important to note that the time-zero data point has been excluded from Figure \ref{kinetix} as the time-resolution of the TRPL (3 ns) is insufficient to resolve the initial picosecond decay of the $S_1$ population (see supplementary materials for optically-gated TRPL experiments). The extracted spectral weightings at time-zero has $S_1>$ME (Figure S8), but within the 3 ns time resolution the ME state becomes dominant. The ratio of ME and $S_1$ weightings are approximately constant for this initial time period ($\lesssim 25$\,ns). After 25\,ns, the ratio of ME to $S_1$ plummets (Figure \ref{kinetix} B), and the $S_1$ spectrum once again becomes dominant after 65\,ns. From 100\,ns, the ratio of ME and $S_1$ weightings is again constant, with both spectra decaying with a long time constant.

We applied several kinetic models in an attempt to reproduce the observed data. In the first model, we constrained $k_3$ and $k_3'$ to be zero, and fixed $k_1$, $k_S$, and $k_T$ from the results of TRPL experiments (Figure S12). The free kinetic parameters are thus $k_1'$, $k_2$, $k_2'$ and $k_{ME}$. Unsurprisingly, this model completely fails to capture the essence of the TRPL. The results are shown as a grey dashed curve in Figure \ref{kinetix} B. Without the direct recombination channel ($k_3$, TTA), the ratio of ME to $S_1$ cannot fall below that initially established with the $S_1 + S_0 \rightleftharpoons ME$ equilibrium.

A second model, which constrains $k_2$ and $k_2'$ to be zero, while allowing direct singlet fission (rate $k_3'$) and TTA ($k_3$), fits the data admirably, as shown by the solid lines in Figure \ref{kinetix} A and B. The initial decay is described by an equilibrium between $S_0$ and ME while the population of $T_1$ increases due to a direct fission mechanism ($k_3'$), while both $S_1$ and ME states decay. After sufficient build-up of the $T_1$ population, it becomes the dominant source of nascent $S_1$ state molecules. This is accompanied by a change in the emission spectrum and the time-constant, now controlled by the $T_1$ lifetime, determined here to be about 60\,$\mu$s. The physicality of this model can be tested by inspecting the ratios $k_1/k_1'$ and $k_3/k_3'$, which result in $\Delta G_1=-0.13$\,eV, and $\Delta G_3=-0.08$\,eV. Both have the appropriate sign, though $\Delta G_3$ is a little smaller in magnitude than expected.

A model which includes both direct and indirect singlet fission naturally also fits the data well, having more free parameters, but is too unconstrained to be meaningful. However, by fixing $\Delta G_3$ to be $-0.08$\,eV, the model does not introduce significant rate constants for the channel whereby the ME excimer dissociates into individual triplets, $k_2$ remaining more than three orders of magnitude below the competing dissociation, $k_1'$. By increasing the magnitude of $\Delta G_3$ towards the expected value (see supplementary materials), the modelling abandons the $ME\rightleftharpoons 2T_1$ channel altogether.

Furthermore, it may be reasoned that direct dissociation of the ME state into triplets is unphysical. The observed decay of the ME state, which controls the first time constant in Figure \ref{kinetix}, has a total rate exceeding $10^8$\,s$^{-1}$. Thus, for the ME state to dissociate into a significant population of triplets, $k_2$ must exceed $\sim10^7$\,s$^{-1}$. However, given that $\Delta G_2$ must be at least 0.2\,eV, the TTA rate constant $k_2'$ would exceed the diffusion limit by several times. As such, it is absolutely clear that the singlet fission (and triplet-triplet annihilation) channel is dominated by a direct mechanism from (to) the $S_1$ state, and that the ME state serves to trap the excited state population.

In films of multicrystalline tetracene, it has been widely assumed that the initial singlet exciton photoluminescence decay is associated with decay to a dull intermediate to singlet fission. At low temperatures this is evidenced by red-shifted excimer-like emission. However, just like the present case, the delayed fluorescence, which is associated with TTA, is dominated by singlet excitons. The behaviour of multicrystalline tetracene is thus consistent with the present findings that the excimer-like state serves as a trap, and is not an intermediate to singlet fission.

The implications for the design of endothermic fission materials are clear. Since the ME excimer has a negligible fission rate, and its population naturally exceeds that of the fissionable $S_1$ state after equilibration, the ME excimer state is undesirable. It traps population and thus attenuates singlet fission, which must then compete against the decay of the ME state. Since cofacial $\pi$-chromophores are liable to exhibit excimer formation, we argue that these cannot be efficient endothermic singlet fission systems. Rather, for efficient endothermic fission to be realised, $\pi-\pi$ interactions must be controlled. Achieving this will enable singlet fission to reach its full potential in enhancing the energy conversion efficiency of photovoltaic devices.

\newpage


\section*{Acknowledgments}
T.W.S. acknowledges the Australian Research Council for a Future Fellowship (FT130100177).  This work was supported by the Australian Research Council Centre of Excellence in Exciton Science (CE170100026).

\section*{Contributions}
CBD and LF performed the measurements. CBD, JKG and LF analyzed the data. AJP synthesized the TIPS-Tc material. JEA designed and provided the TIPS-Tc material. MJC provided the PdPQ$_4$ material. TWS, JKG and LF wrote the manuscript. JKG designed the figures with input from TWS. TWS conceived the experiments.

\section*{Supplementary information}
Materials and Methods\\
Supplementary text\\
Figures. S1 to S14\\
Table. S1

\end{document}